\documentclass{sig-alternate-pre}

\usepackage{algorithm}
\usepackage{algpseudocode}
\usepackage{amsmath}
\usepackage{amssymb}
\usepackage{array}
\usepackage{balance}
\usepackage[pangram]{blindtext}
\usepackage[noadjust]{cite}
\usepackage{caption}
\usepackage{color}
\usepackage{comment}
\usepackage{dblfloatfix}
\usepackage[inline]{enumitem}
\usepackage{epsfig}
\usepackage{etoolbox}
\usepackage{fancybox}
\usepackage{filecontents}
\usepackage{fixltx2e}
\usepackage{float}
\usepackage{graphicx}
\usepackage[bookmarks=false]{hyperref}
\usepackage{hyperref}
\usepackage{lastpage}
\usepackage{lipsum}
\usepackage{listings}
\usepackage{multirow}
\usepackage{multicol}
\usepackage{pifont}
\usepackage{placeins}
\usepackage{rotating}
\usepackage{setspace}
\usepackage{soul}
\usepackage[hang, tight]{subfigure}
\usepackage{threeparttable}
\usepackage{times}
\usepackage{url}
\usepackage{verbatim}
\usepackage{wrapfig}
\usepackage[usenames,dvipsnames]{xcolor}
\usepackage{color}

\algtext*{EndWhile}
\algtext*{EndIf}
\algtext*{EndFor}
\algtext*{EndProcedure}

\newbool{inccomment}
\setbool{inccomment}{true}
\newcommand{\XX}[1]{\ifbool{inccomment}{{\color{magenta} #1}}{}}
\newcommand{\CT}[1]{\ifbool{inccomment}{{\color{magenta}CT\@: #1}}{}}
\newcommand{\NT}[1]{\ifbool{inccomment}{{\color{blue}NT\@: #1}}{}}
\newcommand{\TD}[1]{\ifbool{inccomment}{{\color{orange}#1}}{}}
\newcommand{\FN}[1]{\ifbool{inccomment}{{\color{OliveGreen}#1}}{}}
\newcommand{\GR}[1]{\ifbool{inccomment}{{\color{Tan}#1}}{}}
\newcommand{\LD}{\ifbool{inccomment}{{\color{magenta}\\============================================\\}}}
\newcommand{\RF}{\ifbool{inccomment}{{\color{green}~[R]}}}

\newcommand{\roma}[1]{\uppercase\expandafter{\romannumeral #1\relax}}

\begin{document}
\pdfpageheight 11in
\pdfpagewidth 8.5in
\title{HASP: A High-Performance Adaptive Mobile Security \\ Enhancement Against Malicious Speech Recognition\\
\vspace{-6mm}
}

\author{
\alignauthor
	\large{Zirui Xu$^\dagger$, Fuxun Yu$^\dagger$,Chenchen Liu$^\ddagger$, Xiang Chen$^\dagger$}\\
	\vspace{1.5mm}
        \normalsize{$^\dagger$Department of Electrical and Computer Engineering,
        George Mason University, Fairfax, VA, USA 22030\\
        $^\dagger$\{zxu21, fyu2, xchen26\}@gmu.edu}\\
        \normalsize{$^\ddagger$Department of Electrical and Computer Engineering,
        Clarkson University, Potsdam, NY, USA 13699-5720\\
        $^\ddagger$\{chliu\}@clarkson.edu}
        \vspace{2.5cm}
        \\
}


\maketitle
\vspace{-10mm}

\begin{abstract}
Nowadays, machine learning based \textit{Automatic Speech Recognition} (ASR) technique has widely spread in smartphones, home devices, and public facilities.
	As convenient as this technology can be, a considerable security issue also raises -- the users' speech content might be exposed to malicious ASR monitoring and cause severe privacy leakage.
	In this work, we propose \textit{HASP} -- a high-performance security enhancement approach to solve this security issue on mobile devices.
	Leveraging ASR systems' vulnerability to the adversarial examples, \textit{HASP} is designed to cast human imperceptible adversarial noises to real-time speech and effectively perturb malicious ASR monitoring by increasing the Word Error Rate (WER).
	To enhance the practical performance on mobile devices, \textit{HASP} is also optimized for effective adaptation to the human speech characteristics, environmental noises, and mobile computation scenarios.
	The experiments show that \textit{HASP} can achieve optimal real-time security enhancement: it can lead an average WER of $84.55\%$ for perturbing the malicious ASR monitoring, and the data processing speed is 15$\times$$\sim$ 40$\times$ faster compared to the state-of-the-art methods.
	Moreover, \textit{HASP} can effectively perturb various ASR systems, demonstrating a strong transferability.
\end {abstract}

\vspace{-3mm}
\section{Introduction}
Recent progress in the artificial intelligence especially the machine learning technology is reshaping our life in various aspects.
	The most representative example is the \textit{Automatic Speech Recognition} (ASR).
	Enhanced by deep learning models, ASR now can achieve the recognition performance as accurate as human perception.
	Therefore ASR has been maturely applied to embedded voice assistant~\cite{amazonecho, applesiri} and various online voice services~\cite{bing,google}.

Although this technology offers great convenience, a considerable security issue also raises.
	Since ASR widely spreads in smartphones, home devices, and public facilities~\cite{amazonecho, applesiri}, the users' speech content might be exposed to malicious ASR monitoring and cause severe privacy leakage~\cite{monitoring}.
	As reported in ~\cite{link1}, some ASR equipped smart home devices might monitor the users' daily conversation for activity analysis. Also,~\cite{link2} shows that the malicious ASR monitoring can be used to extract the users' voice command biometrics and defraud the mobile voice authentication.
	As ASR's data processing bandwidth, precision, and application popularity keep involving, the security issue of malicious ASR monitoring also becomes more and more severe.

Although malicious ASR monitoring could cause considerable privacy leakage, the machine learning based ASR itself also has significant vulnerability, especially to the ''adversarial examples''~\cite{adversarial}.
	The ASR adversarial examples here can be referred to the speech data with adversarial noises injected~\cite{audioadversarialexample}.
	The adversarial noises could perturb ASR systems and even manipulate the recognition results by increasing the Word Error Rate (WER).
	While, the speech content with adversarial noises still sounds the same to the human perception~\cite{audioadversarialexample,commandersong}.
	Therefore, many works are also proposed to manipulate the speech data to defect the ASR:
	In~\cite{audioadversarialexample}, Carlini \textit{et al.} injected adversarial noises to a speech data with less than 0.1\% waveform difference. Such a speech data can mislead ASR to any desired results.
	In~\cite{houdini}, Cisse \textit{et al.} also crafted a speech adversarial example, and this example can effectively disturb multiple ASR systems.
	Although, such a vulnerability would significantly defect the desired ASR functionality, it can be also well utilized for preventing malicious ASR monitoring~\cite{commandersong}.

In this work, we propose \textit{HASP} -- a high-performance adaptive security enhancement solution, targeting malicious ASR monitoring on mobile devices.
	Working locally on the mobile device, \textit{HASP} is expected to utilize the adversarial example method to inject adversarial noises directly on raw user speech data collected from the microphone.
	Hence, before the speech data is released to local or online third party applications, the user can use \textit{HASP} to encrypt their speech content preventing malicious ASR monitoring.

As many adversarial example based ASR perturbation works still remained in theoretical evaluation, we take several practical challenges into consideration to design and implement \textit{HASP}:
	(1) The proposed ASR perturbation method should meet the real-time requirement with fast processing speed.
	(2) It should balance the adversarial noise impact and human perception quality.
	(3) It should adapt to practical applications, regarding the external environment and internal system factors, such as computation resources and ASR model variance.

To tackle these challenges, we made the following contributions:

\vspace{-2.2mm}
\begin{itemize}
	\item We first proposed a fast ASR perturbation method targeting the speech transformation stage with Mel-frequency Cepstral Coefficients (MFCCs).
	Adversarial noises are derived from MFCC feature vectors and injected to the speech data to perturb malicious ASR monitoring.
		\vspace{-1.8mm}
	\item We further optimized the perturbation method to reduce the adversarial noises' impact to human perception quality, while the perturbation effectiveness is well maintained.
		\vspace{-1.8mm}
	\item We also enhanced the adaptability of \textit{HASP}, regarding the environmental noises, mobile computation scenarios, and ASR model variation.
		\vspace{-1.8mm}
	\item We implemented the proposed \textit{HASP} as a system component on Android smartphones, and further evaluated the real-time performance in practical utilization scenarios.
\end{itemize}
\vspace{-2mm}

Experimental results show that:
	the proposed MFCC based ASR perturbation method of \textit{HASP} could achieve an average WER of 84.5\% to the most representative ASR model (\textit{i.e.} \textit{DeepSpeech} ~\cite{deepspeech}).
	The speech quality oriented optimization of \textit{HASP} can further reduce the human-perceptible adversarial noises by 42\%.
	Comparing to the state-of-the-art ASR perturbation methods, \textit{HASP} could achieve 40$\times$ faster data processing speed to fulfill the real-time processing requirement.
	Regarding the adaptability, \textit{HASP} can be well deployed with environmental noises as loud as 81\textit{dB}, and support different application scenarios' in real-time.
	Moreover, by evaluating \textit{HASP} with multiple state-of-the-art ASR systems (\textit{i.e.} Google Speech Recognition~\cite{google}, Microsoft Bing Voice Recognition~\cite{bing}, \textit{etc.}), \textit{HASP} also demonstrates significant model transferability with WER from 51\% to 77\%, which indicates strong capability to handle potential unknown ASR systems.

\section{Preliminary}
\label{prelim}
In this section, the technical preliminaries are presented to facilitate our design.
	Firstly, we investigate the ASR processing flow and the significance of the fundamental MFCC process.
	Then, we brief the state-of-art ASR perturbation works.
	Last but not least, we review the speech processing flow on mobile devices to illustrate the malicious ASR monitoring and possible defense strategy.

\begin{figure}[t!]
	\centering
	\captionsetup{justification=centering}
	\includegraphics[width=3.3in]{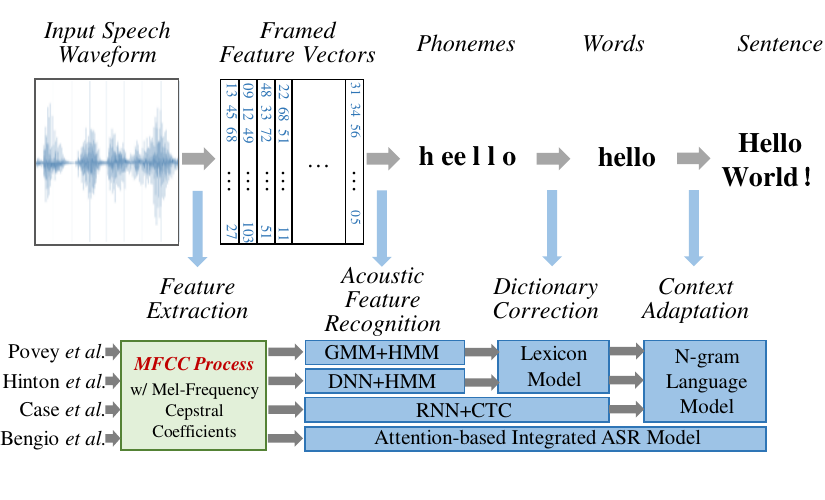}
	\vspace{-8mm}
	\caption{Typical ASR Processing Flow with Different Models}
	\label{ASR_Process}
	\vspace{-2mm}
\end{figure}

\subsection{Automatic Speech Recognition (ASR)}
ASR allows machines to recognize and convert human speech content to text automatically~\cite{yu2016automatic}.
	As shown in Fig.~\ref{ASR_Process}, different ASR systems generally utilize 4 major stages to process the speech data from speech waveform to sentence text~\cite{miao2015eesen}:
	The framed speech waveform is firstly converted to MFCC feature vectors with certain frequency patterns;
	Then, the MFCC feature vectors are recognized by the acoustic models as phonemes with the highest probability, and the combined phonemes are used to estimate potential letter sequence;
	Specific dictionary models are also used to correct the word spelling of the phoneme letter sequence;
	At last, the words are further adjusted to the contexts and merged to the final sentence text by language models.

Through ASR processing, various models are utilized in each stage.
	For acoustic feature recognition, as the features are processed with certain correlations, temporal models are well utilized in addition to further feature extraction, \textit{e.g.} Hidden Markov Models with Deep Neural Network (DNN$+$HMMs) \cite{graves2014towards}.
	Recently, advanced integrated models also emerged targeting multiple stages, \textit{e.g.} Recurrent Neural Network with Connectionist Temporal Classification (RNN$+$CTC)~\cite{sak2015fast} and Attention-based ASR model~\cite{chorowski2015attention}.

Although various models present in ASR systems, MFCC process is constantly adopted in the first feature extraction stage regardless of ASR system variance~\cite{kepuska2015robust}.
	MFCC process transforms the speech waveform to feature vectors composed of a set of short-term power spectrum coefficients, which collectively make up a Mel-Frequency Cepstrum (MFC)~\cite{ahmad2015unique}.
	As MFC approximates the human auditory system's response more closely than other linear cepstrum methods, MFCC feature vector is considered the most suitable frequency transformation format for speech data.
	Comparing to other models in ASR systems, MFCC process not only lays the most primary and intrinsic feature foundation for ASR processing, but also has much less computation overhead.

Considering the significance of MFCC process, we assume that perturbing the MFCC feature vectors could cause more influential impact to the whole ASR system with better computation efficiency.
	In later sections, we further demonstrate the mechanism of MFCC process and propose our ASR perturbation method based on the manipulation of MFCC feature vectors.

\subsection{\fontsize{11}{12}\selectfont \textbf{ASR Perturbation with Adversarial Examples}}
Recently, inspired by the adversarial attacks to the neural networks for image recognition~\cite{adversarial}, many works have been proposed to apply adversarial examples to perturb the ASR process~\cite{audioadversarialexample,didyouhearthat,craftingadversarialexample}.
	The ASR adversarial examples are crafted by injecting adversarial noises to the speech data, which can effectively fool the ASR system but still sound the same to human perception~\cite{audioadversarialexample}.

In~\cite{didyouhearthat}, Alzantot \textit{et al.} crafted several voice commands with trivial background adversarial noises, which were generated with genetic algorithm.
	Although the crafted adversarial noises can effectively mislead ASR to totally different meanings, the generation method can be applied only to specific short voice command with small perturbation scope.
In~\cite{commandersong}, Yuan \textit{et al.} embedded the voice commands into songs stealthily as adversarial noises to activate specific ASR systems.
	Although they investigated the transferability of the adversarial noises to various ASR systems, their method was only transferable to two different systems at most.
	Meanwhile, their method was realized at a considerable cost of computation overhead, which significantly compromised the practical application.
In~\cite{craftingadversarialexample}, Gong \textit{et al.} crafted adversarial noises that can manipulate ASR to desired results. Although highly effective, the generation process also took extremely long computation time.

These works have demonstrated ASR's significant vulnerability to the adversarial examples, which can be well utilized to prevent malicious ASR monitoring.
	However, most of the ASR perturbation works still suffer from obvious issues, such as perturbation scope, model transferability, and computation overhead.
	In this work, our proposed method is expected to have universal effectiveness regardless of ASR model variance and optimal processing speed with low computation overhead.

\begin{figure}[b]
	\centering
	\captionsetup{justification=centering}
	\vspace{-2mm}
	\includegraphics[width=3.3in]{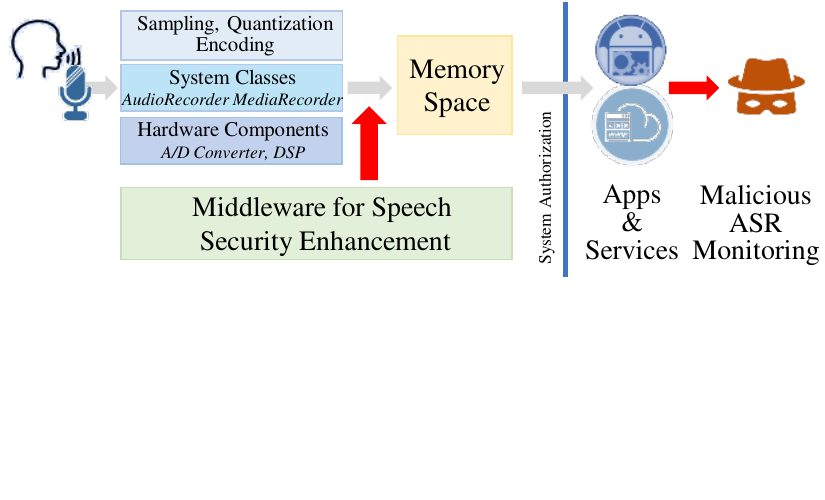}
	\vspace{-27mm}
	\caption{ASR Process Perturbation on Mobile System}
	\label{Audio_Process_in_Mobile}
	\vspace{-2.5mm}
\end{figure}

\subsection{ASR on Mobile System}
Fig.~\ref{Audio_Process_in_Mobile} briefly illustrates the speech processing flow on an Android mobile system~\cite{android}.
	After being collected and amplified by microphone, the speech data is further sampled, quantized, and encoded by on-board hardware components, such as Analog/Digital (A/D) Converter and Digital Signal Processor (DSP).
	The hardware components are controlled by several software classes provided by Android system, such as \textit{AudioRecorder} and \textit{MediaRecorder}~\cite{mediarecorder}.
	The processed speech data is then stored at specific memory space designated by the system before releasing to local applications or online services for further utilization.
	However, malicious ASR is usually embedded in those applications and services.
	As long as the audio authority is granted to these embedded targets by the Android system, malicious ASR can easily monitor the speech content.

In that case, we propose a security enhancement middle-ware before the speech data storage in the memory space.
	This middle-ware can leverage ASR's vulnerability to adversarial examples, and inject adversarial noises to the raw speech data before exposed to malicious ASR monitoring.
	The users can also use the enhanced mobile device as a voice changer to interact with other ASR embedded devices for better security.

In the following sections, we will dive into the design details.

\section{ASR Perturbation based on\\ Adversarial MFCC Noise}
\label{method}
\begin{figure}[b]
	\centering
	\captionsetup{justification=centering}
	\vspace{-5mm}
	\includegraphics[width=3.3in]{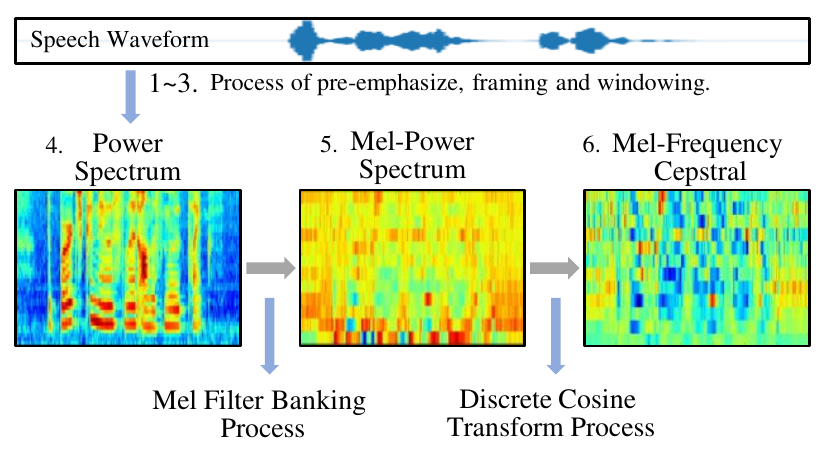}
	\vspace{-6mm}
	\caption{MFCC Process Analysis with Spectrum Examples}
	\label{MFCC_Process}
	\vspace{-1mm}
\end{figure}
In the last section, we introduced the significance of the fundamental MFCC process, which is an efficient and effective manipulation target for perturbing malicious ASR monitoring.
	In this section, we will further investigate MFCC process mechanism and propose an ASR perturbation method based on the adversarial examples derived from the MFCC feature vectors.

\subsection{MFCC Process Analysis}
MFCC process transforms an speech waveform into feature vectors composed of coefficients of Mel-Frequency Cepstrum.
	Such a process usually contains 6 steps. Major steps and the corresponding outputs are demonstrated in Fig.~\ref{MFCC_Process}:

\vspace{1mm}
1. \textit{Pre-emphasize}.
Before MFCC process, assuming an input speech waveform $x$ is pre-processed by sampling and quantization, it is firstly transformed into a speech vector $x_s$ in time domain, where $s$ indicates the $s_{th}$ sampling point~\cite{muda2010voice}.

The pre-emphaize process is used to balance the frequency spectrum of $x_s$, especially on the weak high-frequency domain that needs amplification.
	To achieve the spectrum balance, the pre-emphasis filter can be applied to  $x_s$ with the following equation:

\vspace{-1mm}
\small
\begin{equation}
	PEM:	y^{pre}_s=x_s - \alpha x_{s-1},
	\label{eq:eana}
	\vspace{-0.5mm}
\end{equation}
\normalsize
where $\alpha$ is the filter coefficient ($\sim$0.97), and $y^{pre}_s$ is the balanced speech vector after the pre-emphasis process.

\vspace{1mm}
2. \textit{Framing}.
	The pre-emphasized $y^{pre}_s$ needs to be partitioned into multiple vector frames, since Fourier-transform can't be applied across the entire speech without continuous frequency contour loss through time domain~\cite{gaikwad2010review}.
	In Fourier transform of speech, frequencies are assumed to be relatively stationary in a very short time frame. Therefore, by applying the Fourier transform to each frame, we can obtain a good approximation of the frequency contours, and the entire audio can be presented by concatenating adjacent frames together.
	The frame size is usually set as 20\textit{ms}$\sim$40\textit{ms}, and $50\%$ overlap is reserved between consecutive vector frames~\cite{gaikwad2010review}.

After the framing process, the speech vector is converted into a vector frame array, where the $m_{th}$ point in the $n_{th}$ vector frame is represented as $y^{fra}_{(m,n)}$.

\vspace{1mm}
3. \textit{Windowing}.
As certain overlap present between vector frames, the vector frames can't be well aligned with smooth connection, which may lead to considerable frequency leakage in frequency domain.
	To precisely align the vector frames, a hamming-windowing function is to polish each signal point $y^{fra}_{(m,n)}$ to $y^{win}_{(m,n)}$~\cite{allen1977short}:

	\vspace{-2mm}
\small
\begin{equation}
	\vspace{-2mm}
	WIN:	y^{win}_{(m,n)}=\left\{ 0.54 -0.46 cos (\frac{2 \pi (n-1)}{N^{fra}-1})\right\} \times y^{fra}_{(m,n)},
	\vspace{-0.5mm}
	\label{eq:eana}
\end{equation}
\normalsize
where $N^{fra}$ means the windowing length in each frame.

4. \textit{Fast Fourier-Transforming} (FFT).
Since it is difficult to extract features from the speech vectors in time domain, FFT is widely utilized to transfer the speech vectors into frequency domain and derive their power spectrums~\cite{moskowitz1964estimates}.

	To achieve frequency domain transformation of $y^{win}_{(m,n)}$, we first apply $N^{FFT}$-point FFT to each vector frame:

\vspace{-3mm}
\small
\begin{equation}
	\medmuskip=-5mu
	\vspace{1mm}
	FFT:	y^{FFT}_{(k,n)}=\sum_{n=1}^{N^{FFT}} y^{win}_{(m,n)} e^{-j2 \pi k n /N^{FFT}}, \quad 1 \le k \le K,
	\label{eq:e}
	\vspace{-2mm}
\end{equation}
\normalsize
where $y^{FFT}_{(k,n)}$ is the frequency spectrum of the $n_{th}$ vector frame, and $k$ means the $k_{th}$ frequency point from $K$ points in total.

In frequency domain, $y^{FFT}_{(k,n)}$ is further transformed to power spectrum for feature extraction:

\small
\begin{equation}
	y^{FFT}_{(k,n)}=\frac{1}{N^{FFT}}|{y^{FFT}_{(k,n)}}^2|.
	\label{eq:ee}
\end{equation}
\normalsize

	As shown in in Fig.~\ref{MFCC_Process}, the power spectrum distributions demonstrate obvious frequency patterns.

\vspace{1.5mm}
5. \textit{Mel-Filter Banking}.
In MFCC process, Mel-Filter Banking process is adopted to highlight the frequency patterns regarding human perception preference~\cite{muda2010voice}.
	It is composed of a set of $L$ filters ($L$ =13$\sim$26), which are more discriminative to low frequencies but less discriminative to higher ones.
	By applying Mel-Filter Bank of \textit{MB}$(f_1, f_2,...,f_L)$, a Mel-power spectrum $y^{Mel}_{ln}$ can be derived as:

\vspace{-1mm}
\small
\begin{equation}
MFB:	y^{Mel}_{(l,n)}=P^{FFT}_{kn} \times MB(f_1, f_2,...,f_L),\quad 1 \le l \le L,
	\label{eq:ea}
\end{equation}
\normalsize
where $l$ is the $l_{th}$ point in the $n_{th}$ frame, and $L$ also indicates the total point number in each frame after Mel-Filter Banking process.

As shown in Fig.~\ref{MFCC_Process}, although Mel-power spectrum has less resolution than the original power spectrum, its feature complexity in the low frequency domain is well enhanced.

\vspace{1.5mm}
6. \textit{Discrete Cosine Transform} (\textit{DCT}).
Mel-power spectrum usually has highly correlation inside, which could be problematic for machine learning algorithms.
	Hence, to accommodate later machine learnig models in ASR, dedicated DCT process is applied to decorrelate the $y^{Mel}_{(l,n)}$:

\vspace{-3mm}
\begin{equation}
	DCT:	y^{DCT}_{(l,n)}=y^{Mel}_{ln} cos[(l-0.5) \frac{\pi l}{L}].
	\label{eq:eaa}
	\vspace{-1mm}
\end{equation}

As the MFC patterns in Fig.~\ref{MFCC_Process} become more granular, the effectiveness of such a decorrelation process can be well observed.

\vspace{1.5mm}
With the aforementioned 6 steps, MFCC process transforms a speech waveform $x_s$ to a set of MFCC feature vectors of $y^{DCT}_{(l,n)}$, which offers distinct speech perceptive features for ASR process.

\subsection{Adversarial MFCC Noise Generation}
As aforementioned, MFCC process is an efficient and effective target to perturb ASR.
	Based on the MFCC feature vectors, we propose a novel speech adversarial example generation algorithm, which derives the adversarial noises from MFCC feature vectors.

\subsubsection{Generation Algorithm Design}
\begin{figure}[tb]
	\centering
	\captionsetup{justification=centering}
	\includegraphics[width=3.3in]{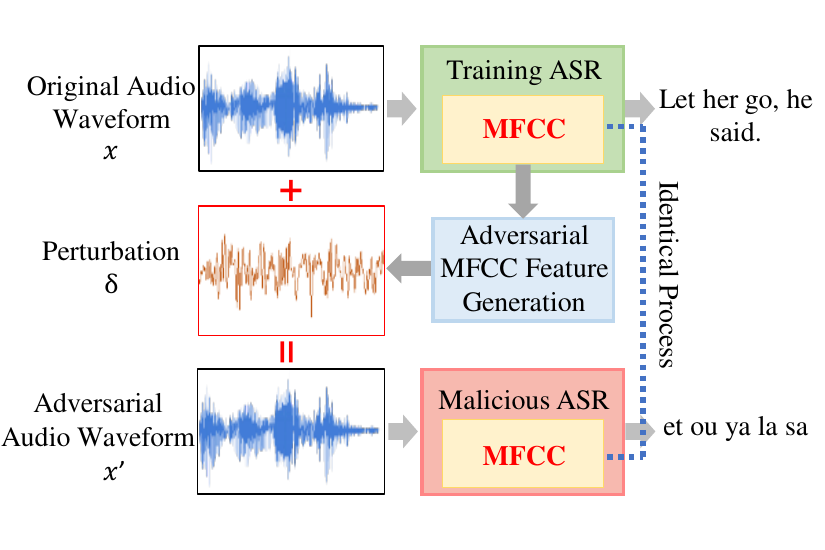}
	\vspace{-8mm}
	\caption{Adversarial Example Process}
	\label{Adversarial_Example_Process}
	\vspace{-4mm}
\end{figure}
The adversarial example generation method is originally proposed in~\cite{adversarial}, targeting image recognition manipulation with adversarial noises in pixels.
	When applied to ASR, the perturbation process shares the same methodology.
	As shown in the left of Fig.~\ref{Adversarial_Example_Process}:
	for any given input speech waveform $x$ in a particular time frame, a small adversarial noises -- $\delta$ can be superposed to $x$ in the frequency domain, and an adversarial example of speech waveform $x'$ is crafted.
	Although $x'$ sounds the same to human perception, it can effectively perturb ASR results.

\textit{Adversarial Example Generation}:
As shown in the right of Fig.~\ref{Adversarial_Example_Process}, our adversarial example generation method is targeting MFCC process, considering its small computation cost and consistency across various ASR systems.

To further reduce the computation overhead, we choose Fast Gradient Method (FGM) as the fundamental method to generate the perturbation noise from MFCC feature vectors $y^{DCT}_{(l,n)}$~\cite{adversarial}.

The general FGM can be formulated as:

\vspace{-1mm}
\small
\begin{equation}
	\delta= \bigtriangledown J(\theta, X, Y),
	\label{eq:fgm}
	\vspace{-0.5mm}
\end{equation}
\normalsize
where $X$ represents the input to the target model of $M$ with parameter configuration of $\theta$, while $Y$ indicates the desired manipulation results.
	Let $f(\cdot)$ represents the function of $M$. 
	$J$ is the cost function that measures the difference between $f(X+\delta)$ and $Y$, and $\bigtriangledown$ is a partial differentiate process.

By partially differentiating the cost function $J$ iteratively, $f(X+\delta)$ will gradually approach to $Y$.
	When $Y=f(X+\delta)$, $\delta$ can be considered as effective adversarial noise (or perturbation) that can manipulate the classification result of $M$.

\vspace{1mm}
\textit{Adversarial MFCC Noise}:
In our proposed ASR perturbation methods, we apply FGM to MFCC process for the adversarial MFCC noise generation by replacing:
	$M$ and $\theta$ with MFCC transformation;
	$X$ with MFCC feature vectors -- $y^{DCT}_{(l,n)}$;
	and set $Y$ to 0 for maximizing the perturbation effectiveness regardless the speech content.
	To further fulfill Eq. 7 with MFCC, we replace $y^{DCT}_{(l,n)}$ with its calculation process through Eq.~\ref{eq:eana} to Eq.~\ref{eq:eaa}, which transforms a speech waveform of $x_s$ to MFCC feature vectors:

\vspace{-3mm}
\small
\begin{equation}
	\medmuskip=-10mu
	\delta^{MFCC}= \bigtriangledown J(\theta^{MFCC}, DCT(MFB(FFT(WIN(PEM(x_s)))), 0).
	\label{eq:fgm2}
	\vspace{0mm}
\end{equation}
\normalsize
.

According to the chain rule of partial differentiation, we differentiate $J$ from Eq.~\ref{eq:eaa} to Eq.~\ref{eq:eana} of MFCC process step by step. Such a differentiation can be formulated as:

\vspace{-2mm}
\small
\begin{equation}
	\medmuskip=-1mu
	\begin{aligned}
		\delta^{MFCC}_{(l,n)}= & DCT'(\cdot)\cdot MFB'(\cdot) \cdot FFT'(\cdot) \cdot WIN'(\cdot) \cdot PEM'(\cdot), \\
		& (\quad x_m + \delta^{MFCC}_{(l,n)} \to 0 \quad and \quad \delta^{MFCC}_{(l,n)}<T_{adv}),
		\label{eq:dif1}
	\end{aligned}
	\vspace{-1mm}
\end{equation}
\normalsize
As the derivation value of Eq.~\ref{eq:eana} and Eq.~\ref{eq:ea} are constants, which can be obtained directly, Eq. 9 can be further simplified as:

\vspace{-2mm}
\small
\begin{equation}
	\medmuskip=-1mu
	\begin{aligned}
	\delta^{MFCC}_{(l,n)}= &MB(f_1, f_2,...,f_L) \times \left\{ 0.54 -0.46 cos (\frac{2 \pi (n-1)}{N^{fra}-1})\right\}  \\
	&\times DCT'(\cdot)\cdot FFT'(\cdot) \cdot PEM'(\cdot), \\
	& (\quad x_m + \delta^{MFCC}_{(l,n)} \to 0 \quad and \quad \delta^{MFCC}_{(l,n)}<T_{adv}).
	\label{eq:dif}
\end{aligned}
	\vspace{-1mm}
\end{equation}
\normalsize
where $\delta^{MFCC}_{(l,n)}$ is the generated adversarial MFCC noise that corresponds to each MFCC feature vector frame.

\vspace{2mm}
From Eq. 10, we can see that as the adversarial MFCC noise is based on vector frame, the eventual MFCC based adversarial example is extremely scalable.
	By concatenating multiple $\delta^{MFCC}_{(l,n)}$ together, we can acquire an adversarial example targeting arbitrary speech length.
	Also, the major computation load in Eq. 10 mainly depends on the derivation of DCT, FFT, and PEM.
	As these calculation are well supported by state-of-the-art computing systems, we can expect an optimal processing speed of the proposed method.
	Moreover, since the generated adversarial MFCC noise is expected to perturbing malicious ASR monitoring while maintain normal content to normal human listeners, we also add a threshold $T_{adv}$ to regulate the generated adversarial MFCC noise from impacting human perception quality.

\subsubsection{Adversarial MFCC Example}
\begin{figure}[b]
	\centering
	\captionsetup{justification=centering}
	\includegraphics[width=3.35in]{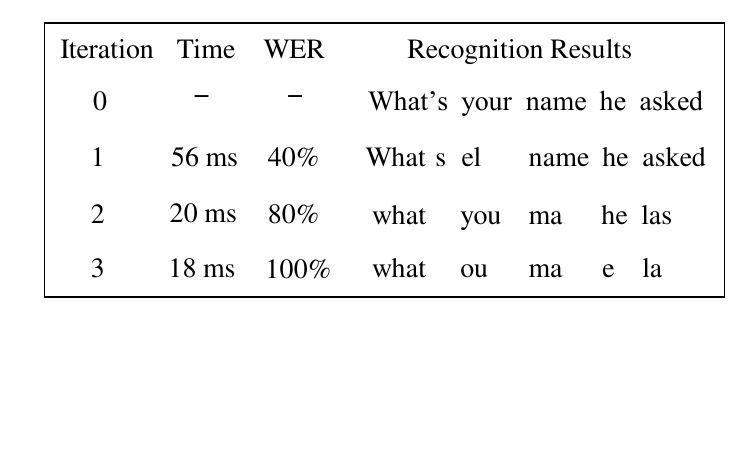}
	\vspace{-18.5mm}
	\caption{A Case Study: Adversarial MFCC Noise Effectiveness}
	\label{Sample_Case}
	\vspace{-1mm}
\end{figure}
Fig.~\ref{Sample_Case} shows a case study for the effectiveness of one adversarial MFCC example.
	From the figure we can see that, the original speech data can be well recognized by ASR system as "What's your name he asked".
	However, with adversarial MFCC noise injecting into the original speech data, the recognition results are gradually defected, and words can't be correctly recognized by ASR system.
	Such a situation is considered as ASR perturbation, which can be evaluated with Word Error Rate (WER).

It is important to note that, the effectiveness of the adversarial MFCC noise is depending on the number of the computation iterations at a cost of the computation time overhead.
	With increasing iterations, the WER of the ASR results could be manipulated to $100\%$.
	In later sections, we will further propose relative optimization methods and evaluations discussing the computation overhead of the proposed method.

\subsection{Optimization Regarding\\ Human Perception Quality}
Although we regulate the noises to reduce the human perception impact during the adversarial MFCC noise generation, the threshold based regulation still lack expected adaptability to the human hearing, and may defect the perturbation effectiveness.
	In this part, we explore the optimization scheme to enhance the proposed method \textit{w.r.t} balancing the adversarial noises and human perception quality.

\subsubsection{Noise Masking Effect Analysis}
\begin{figure}[t]
	\centering
	\captionsetup{justification=centering}
	\includegraphics[width=3.22in]{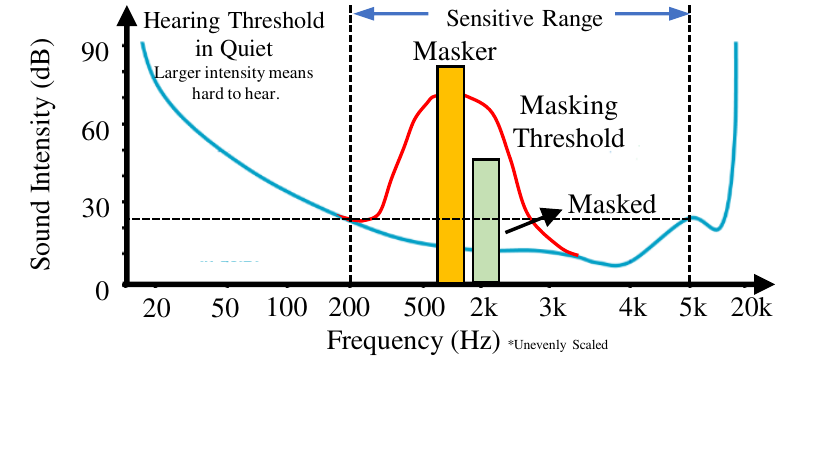}
	\vspace{-10mm}
	\caption{Noise Masking Effect Analysis}
	\label{Masking_Effect}
	\vspace{-2mm}
\end{figure}

The major optimization principle is to hide the adversarial noises form the human perceptive range.
	Therefore, two auditory masking effects can be well utilized to mask the adversarial noises~\cite{gelfand2017hearing}:

(1) \textit{Noise Masking with Frequency Sensitivity}.
As human perceptible frequency ranges from 20\textit{Hz}$\sim$20k\textit{Hz}, significant sensitivity preference inevitably present.
	The hearing sensitivity threshold is shown as the blue line in Fig.~\ref{Masking_Effect}.
	We can see that the sensitive range of hearing perception is between 200\textit{Hz} to 5\textit{kHz} with sound intensity less than 30\textit{dB}~\cite{barnea1990tinnitus}.
	Based on this masking effect, more adversarial noises should be injected to the insensitive range.

(2) \textit{Noise Masking by Adjacent Loudness.}
	The adjacent loudness masking means that the frequency component with higher sound intensity may prevent its adjacent frequency component from human perception.
	As shown in Fig.~\ref{Masking_Effect}, the yellow frequency component creates a masking effect around itself.
	Under the masking threshold shown as the red line, the blue frequency component can be hardly perceived by human hearing.
	However, such a masking effect won't significantly impact ASR process~\cite{gelfand2017hearing}.
	Hence, by identifying the loudness in a speech waveform, more aggressive adversarial noises can be injected around it.

\subsubsection{Optimization Method Design}
	In this work, we integrated two principles into our adversarial MFCC noise generation for method optimization:

According the first masking effect, we first set the threshold of the adversarial noise generation as: $T_{adv}\in\{200Hz,5kHz\}$ in Eq. 9.
	Such a regulation could effectively prevent significant adversarial noise perception.
	However, the limited perturbation range will led to reduced adversarial noise distribution and defect the ASR perturbation effectiveness.

Therefore, we introduce an adversarial noise compensation scheme following the second masking effect.
	For each speech waveform, we locate the top \textit{t}\% frequency component with the highest sound intensity (empirically, \textit{t}$\approx$10).
	Then, around those frequency component, we further regulate the range of $\delta^{MFCC}_{(l,n)}$ to generate concentrated adversarial noises.

Combining these two optimization schemes, we can effectively enhance the adversarial MFCC noise without defecting, but improving the human perception quality instead.

\subsubsection{Optimization Effectiveness}
\begin{figure}[t]
	\centering
	\captionsetup{justification=centering}
	\includegraphics[width=3.3in]{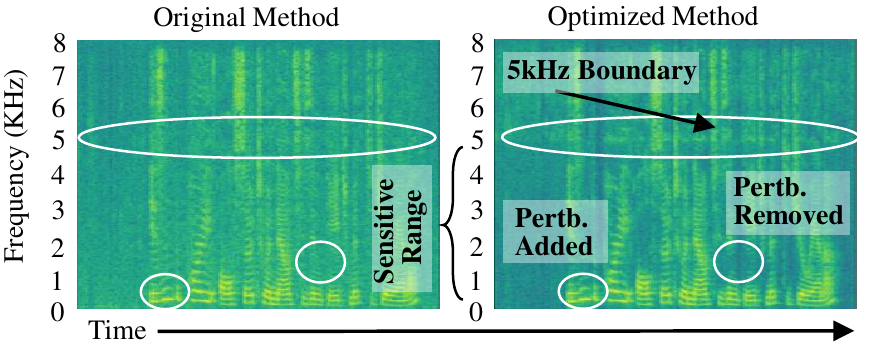}
	\vspace{-2.5mm}
	\caption{Spectrum Comparison for Optimization Demonstration}
	\label{Optimized_Sample_Case}
	\vspace{-1.5mm}
\end{figure}

Two adversarial MFCC example spectrums of are compared in compares Fig.~\ref{Optimized_Sample_Case}.
	The left one is generated with the proposed MFCC feature-based ASR perturbation method as presented in Sec 3.2, while the right one is generated by the optimized method.

Comparing the two specturms, we can find several differences:
	First of all, in the optimized spectrum, a clear boundary with less frequency distribution presents around 5k\textit{Hz} (indicated by lighter color change).
	This is caused by the frequency threshold regulation in the optimization methods, which reduces the adversarial noises within the human perceptible range.
	Moreover, we can see that in the perceptible frequency range, some perturbation is removed, which also reduces the adversarial noises' impact on the manipulated speech data.
	From the figure, we can also find some enhanced perturbation in certain yellow spot (indicated by brighter color change).
	As the yellow spot indicates the loud frequency component from human speech, more adversarial noises are injected around this spot for better perturbation.

Therefore the spectrum differences in Fig.~\ref{Optimized_Sample_Case} further testifies the effectiveness of the proposed method.

\section{PERFORMANCE ENHANCEMENT \\WITH ADAPTIVE SCHEMES}
\label{sys}
In last section, we proposed MFCC feature based ASR perturbation method and further optimize it according to the human perception quality.
In this section, by taking the practical application challenges into consideration, we further enhanced the proposed method in three aspects, \textit{w.r.t} environmental noise compatibility, computation scenario adaptability, and ASR model transferability.

\subsection{Environmental Noise Compatibility}
In practical ASR utilization scenario (\textit{esp.} with mobile devices), the input speech waveform is usually collected with inevitable environmental noises involved.
Therefore the proposed ASR perturbation method requires certain environmental noise compatibility to adapt to the practical scenarios.

When taking the environmental noises into consideration, we formulate it as $\delta^{ENV}$ -- an additional component to the adversarial MFCC noise.
	When applied $\delta^{ENV}$ to Eq. 7, we can derive:

\vspace{-1mm}
\small
\begin{equation}
	\delta= \bigtriangledown J(\theta, X+\delta ^{ENV}, Y),
	\label{eq:env}
	\vspace{-0.5mm}
\end{equation}
\normalsize
where the environmental noise actually contribute to the adversarial MFCC noise generation as $(x+\delta ^{ENV}+ \delta)$.

In such a case, we can well leverage the contribution of the environmental noises to reduce the adversarial noise generation load and therefore enhance the process efficacy.
\begin{table}[b]
	\centering
	\captionsetup{justification=centering}
	\vspace{-2mm}
	\caption{Environmental Noise and Perturbation Iteration Relationship}
	\vspace{-3mm}
	\includegraphics[width=3.3in]{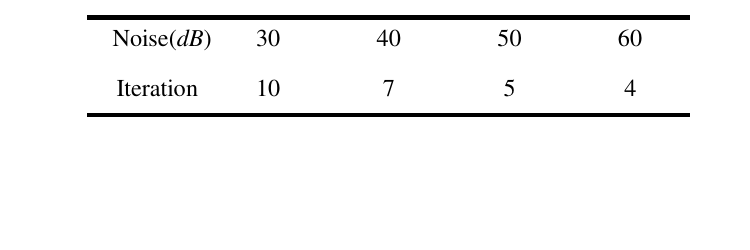}
	\vspace{-14mm}
	\label{Relation_Noise_Iteration}
	\vspace{-5mm}
\end{table}
	Tab.~\ref{Relation_Noise_Iteration} shows the relationship between the environmental noise intensity and  the computation iterations in terms of process efficiency.
	From the table we can see that, given the same WER of $80\%$, with increment of the environmental noise intensity, the iteration load of the perturbs process is gradually relieved.

However, the environmental noise is not always positively correlated with the adversarial MFCC noise.
	When the environmental noise intensity exceed the maximum value in designated range, Eq.~\ref{eq:env} will become invalid. For example, when $\delta ^{ENV}+\delta$ is larger than $X$, which means the adversarial perturbation noise overwhelm the overall human speech.

Empirically, we tested our method in several environments with different noise intensity, such as room, office, street, and supermarket.
	According to the testing, the maximum tolerance value of environmental noise intensity is 81\textit{dB}.

\subsection{Computation Scenario Adaptability}
Besides the environmental noises, different computation scenarios also raise significant real-time requirements.
	As aforementioned in Section 2.3, the proposed ASR perturbation method can be also used as a real-time speech encryptor (\textit{e.g.} a voice changer).
	To prevent malicious ASR monitoring while maintaining certain user experience quality, our proposed method also need sufficient enhancement to adapt real-time computation scenarios.

The major real-time issue in speech processing is the time delay.
	And different mobile applications may have various delay tolerance due to the processing mechanism and utilization scenarios.
	For example, in the telephone calling scenario, where all the speech data is required to be transmitted in real-time, the delay tolerance is required to be less than 450\textit{ms}.
	While, in the voice messaging scenario, the delay tolerance is relieved to 1000\textit{ms} due to expected 3rd party server processing.
When we take the adversarial MFCC noise generation into consideration, the real-time delay equals to the computation time after speech.
As the MFCC process is based on vector frames, the proposed method has naturally optimal scalability.
	When the real-time requirement has extremely small delay tolerance, the adversarial MFCC noise generation can be reconfigured with smaller or discrete vector frames to reduce the computation time for each iteration. So with small perturbation effectiveness compromised, the proposed method can still catch up the real-time speech process.
	Meanwhile for high time delay tolerance, we can extend the vector frame scale even to a whole sentence.
	With more comprehensive information and computation time, the adversarial MFCC noise generation can offer significant perturbation effectiveness.

\begin{figure}[b]
	\centering
	\captionsetup{justification=centering}
	\includegraphics[width=3.3in]{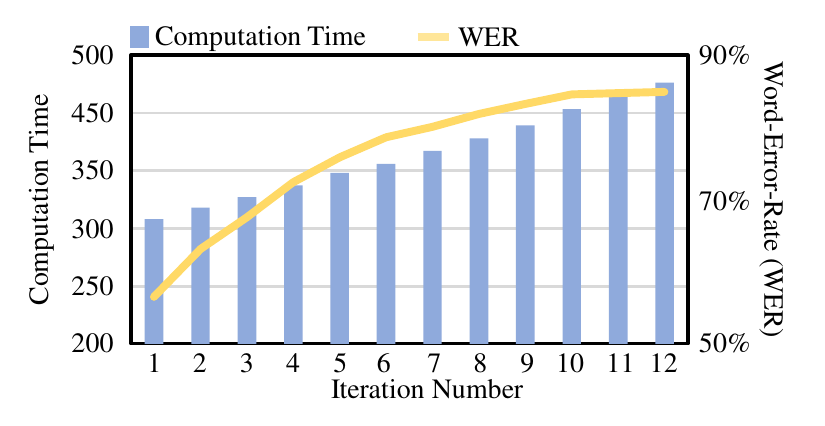}
	\vspace{-1mm}
	\caption{Computation time and WER Relationship}
	\label{Relation_WER_Time}
	\vspace{1mm}
\end{figure}

Considering different delay time tolerance, we test our proposed method on 500 audio samples from Common Voice Dataset~\cite{commonvoice} and Fig.~\ref{Relation_WER_Time} depicts the relationship between iteration, computation time and WER. During the test, adversarial MFCC noise generation method produce adversarial MFCC noise on each 200\textit{ms} length speech data.
In the figure, it clearly shows that with the computation time increase, the WER will also get a higher value.

\subsection{ASR Model Transferability}
As the proposed method is for preventing malicious ASR monitoring. The generated adversarial MFCC noise is expected to encounter different ASR models.
	Therefore the model transferability is most critical key of actual utilization.

As aforementioned in Section 2, the MFCC process is consistently adopted as the very first feature extraction stage in various ASR system.
	Meanwhile MFCC feature vectors feed all the other models in a ASR system.
	Therefore, MFCC process lays the most primary and intrinsic feature foundation for ASR process.
	In such a case, manipulation on the MFCC feature vectors has the natural advantages of model transferability.
	To enhance the transferability of the proposed methods, we only apply the MFCC feature based ASR perturbation method only on the standard MFCC process for adversarial MFCC noise generation in ASR regardless other additional models.
	In later sections, we will further evaluate the ASR model transferability of the proposed method.

\section{SPEECH SECURITY SOLUTION OF HASP}
\label{sys}
Based on the works in previous sections, we proposed \textit{HASP} -- an adaptive mobile speech security enhancement solution.
	The primary method of \textit{HASP} applies the adversarial example generation methodology to the most fundamental ASR feature extraction stage of MFCC.
	The generated adversarial MFCC noises are injected to speech data for perturbing malicious ASR monitoring.
	Based on this fundamental method, we also investigated one intrinsic algorithm optimization targeting improving adversarial MFCC noise's human perception quality, as well as three external capability enhancement schemes regarding environmental noise compatibility, computation scenario adaptability, and model transferability.
	To achieve \textit{HASP} effectively, especially for mobile systems, where are the hot-spot for malicious ASR monitoring, we further discussed the preliminary implementation for it.

	To apply \textit{HASP} to practical utilization, especially the mobile devices, where malicious ASR monitoring prevails, we further implemente \textit{HASP} on Android smartphones.
	To implement the \textit{HASP} in an Android system, we experienced the following major stages:

1. We first used a state-of-the-art smartphone, Google Nexus 5, as our experiment hardware platform.

2. We extracted a standard MFCC model from the \textit{DeepSpeech}~\cite{deepspeech} and modified the MFCC model to support differentiation operation on each stage.

3. We deployed the MFCC feature based adversarial example generation process with Tensorflow library~\cite{tensorflow}, and patched the generation process to be light-weight processing engine.

4. We integrated the native Android DCT and FFT tools in the kernel support the adversarial MFCC noise generation for faster processing speed.

6. We explored the Android kernel programming to the devolve the audio process thread and deployed \textit{HASP} as a middle-ware in the mobile audio processing system.

7. We also optimized the Android system threats management which can make sure the \textit{HASP} can get more system resource after collecting the speech data.

Based on these steps, \textit{HASP} can be well-deployed on the state-of-the-art smartphones with optimal processing performance. In the next section, we will further evaluate \textit{HASP}.

\section{Experiment and Evaluation}
In previous sections, through the design and verification of \textit{HASP}, we have achieved preliminary evaluation of the certain \textit{HASP} performance.
	For example, the speech quality oriented optimization of \textit{HASP} can further reduce the human-perceptible adversarial noises by 42\%, and be well deployed with environmental noises as loud as 81\textit{db}.
	Therefore in this section, we will mainly focus several major performance perspectives, namely the computation performance, perturbation successful rate in terms of WER, as well as ASR model transferability.
	
	In the first experiment, we evaluate the computation performance and perturbation rate by comparing \textit{HASP} to the state-of-art speech manipulation work proposed ~\cite{audioadversarialexample}. \cite{audioadversarialexample} also utilized the similar adversarial example approach to manipulate the ASR results.
	Different from \textit{HASP}, ~\cite{audioadversarialexample} applied the adversarial examples mainly to the CTC model as shown in Fig. 1.
	During the comparison, we adopted audio samples from Common Voice as our dataset~\cite{commonvoice} and use \textit{DeepSpeech}~\cite{deepspeech} as the perturbation targeted malicious monitoring ASR model.

	In the second experiment, we also evaluate \textit{HASP} regarding its model transferability. And multiple commercialized ASR models are also included as malicious monitoring ASR models. Please note that, most of these models are not open-sourced, so this model transferability test also demonstrates \textit{HASP}'s performance on the so-called ``black-box'' ASR models.
	


\subsection{HASP Performance Evaluation}
We first compare the performance between the fundamental method of \textit{HASP}, namely the MFCC feature based ASR perturbation method (refer as "\textit{HASP}"), and the state-of-the-art method proposed in ~\cite{audioadversarialexample} (refered as "CTC").
	As both methods are based on adversarial noise generation, we also introduced random noises as the evaluation baseline (referred as ``Noise'').

Tab.~\ref{Experiment1} shows the comparison results between \textit{HASP}, CTC and Noise in respect of WER and computation time.
	For performance comparison, all the three methods are executed on a computer server to generate adversarial examples over speech data of 1000\textit{mm.}
	
	(1) The left part of Tab. 2 evaluates the perturbation performance by comparing the WER achieved by certain amount of iteration.
	With the same iteration number increment from 1 to 10, 
	\textit{HASP} could achieve a WER ranging from 56.5\% to 84.6\%, 
	and CTC achieves 36.0\% to 77.1\%, while random Noise could also cause certain WER with a rate round 30\%.
	Comparing \textit{HASP} to CTC, we can see that, \textit{HASP} overperforms CTC in every iteration stage.
		
	(2) The right part of Tab. 2 evaluates the computation time performance with certain iteration number.
	From the table we can see that from 1$\sim$10 iterations, the time consumed by \textit{HASP} is maintained around 100\textit{s}.
	Meanwhile, CTC suffers from significant time consumption from 1120\textit{ms} to 3155\textit{ms}.
	In such a case, \textit{HASP} 's processing speed is 15$\times$ $\sim$ 40$\times$ faster then the state-of-art method.

	From this comparison we can see that by deriving the adversarial noises from MFCC feature, we can achieve outstanding efficiency and effectiveness for perturbing malicious ASR process.
	Moreover, for practical ASR perturbation, \textit{HASP} could averagely finish 10 iteration cycles regarding real-time requirement and achieve an average WER of 84.5\%.

\begin{figure}[t]
	\centering
	\captionsetup{justification=centering}
	\includegraphics[width=3.3in]{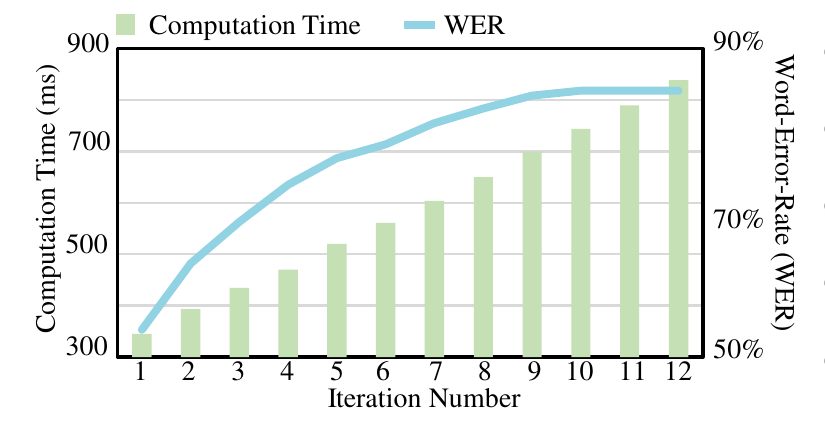}
	\vspace{-5mm}
	\caption{\textit{HASP} Performance Evaluation on Nexus 5}
	\label{Experiment12}
	\vspace{-2mm}
\end{figure}

As aforementioned, we also implemented \textit{HSAP} on the Android smartphone.
Fig.~\ref{Experiment12} illustrates the WER and computation time of \textit{HSAP}.
When deployed on Nexus 5, The adversarial example can be effectively generated for every 200\textit{ms} of speech data.
Regarding the computation time, the time overhead for the first iteration is around 302\textit{ms}. After that,  the average computation cost for each iteration is 42\textit{ms}.


\begin{table}[t]
	\centering
	\captionsetup{justification=centering}
	\vspace{-2mm}
	\caption{WER and Computation Time Performance Comparison on Desktop}
	\vspace{-2mm}
	\includegraphics[width=3.4in]{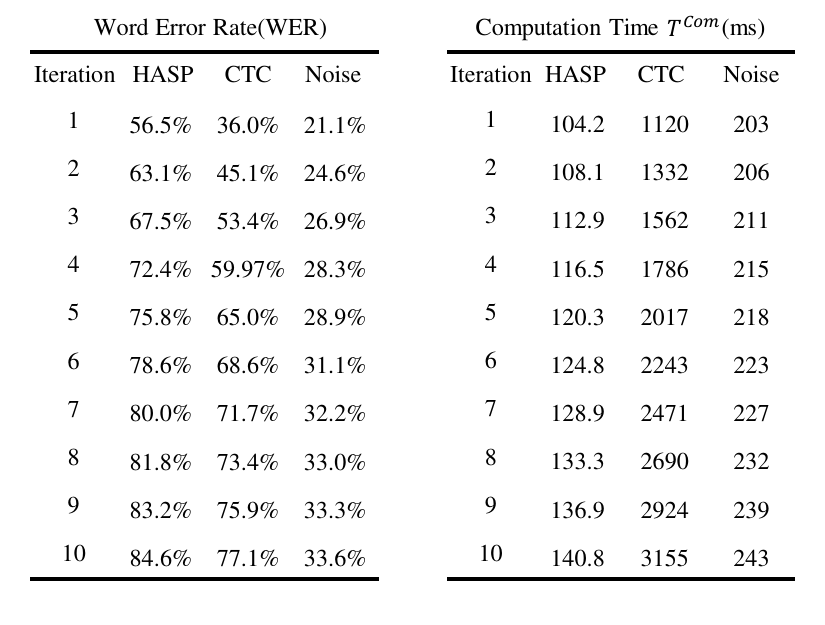}
	\vspace{-5mm}
	\label{Experiment1}
	\vspace{-2mm}
\end{table}

\subsection{Model Transferability of HASP}
\begin{figure}[b]
	\centering
	\captionsetup{justification=centering}
	\includegraphics[width=3.3in]{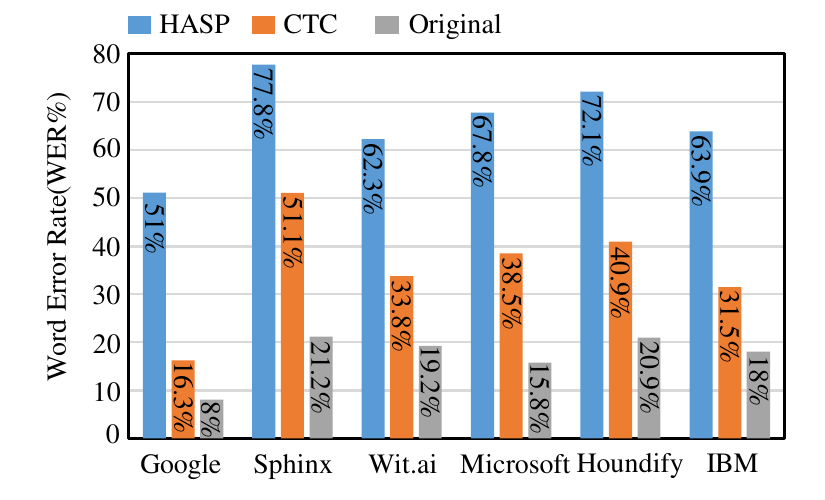}
	\vspace{-5mm}
	\caption{Transferability Evaluation}
	\label{Experiment5}
	\vspace{-2mm}
\end{figure}

In first experiment, we use \textit{DeepSpeech} as our perturbation ASR system. In this experiment, we try to prove that our proposed security enhancement solution has strong transferability that can also perturb other ASR platforms. We choose 6 the state-of-the-art speech recognition platforms: Google Voice, Sphinx, Wit.ai, Microsoft, Houndify and IBM and then generate 500 adversarial examples which are manipulated by \textit{HSAP} and CTC method respectively. Next, we feed these adversarial example audios to 6 ASR systems. Also, in order to make further comparison, we feed 500 original audio examples to the 6 ASR system. Fig.~\ref{Experiment5} illustrates the experiment result: 

In figure, it clearly shows that \textit{HSAP} can always achieve highest WER performance among 6 ASR systems when compared to the CTC perturbed audio and original audio. For \textit{HSAP}, the highest WER value is 77.8\% and obtained when perturb to the Sphinx while the lowest WER value is 51\% and obtained when perturb to the Google Voice. Although testing on Google Voice get the lowest perturbation performance, the WER is more than 50\% which still can protect the speech data from the ASR system. On the other hand, the highest WER which achieved by CTC is 51\% while its lowest WER is 16.3\%. Hence, the experiment result can highly prove that MFCC feature has intrinsic character and universal effectiveness to the all ASR systems and our approach have strong ASR model transferability.

\section{Conclusion}
\label{con}
In this work, we proposed \textit{HASP} -- a high-performance adaptive mobile security enhancement solution targeting malicious ASR monitoring on mobile devices.
	Based on injecting adversarial noises to MFCC feature vectors in ASR, \textit{HASP} achieved outstanding performance in perturbing malicious ASR monitoring, with an average WER of 84.5\% and 15$\times$$\sim$40$\times$ faster processing speed comparing to the state-of-the-art method.
	\textit{HASP} also had high robustness to external noises and application-oriented computation scenarios.
	Moreover, \textit{HASP} demonstrated optimal perturbation transferability to different ASR systems.
	Therefore \textit{HASP} can be well deployed on mobile devices and meet practical usage requirements.
	Through the design and implementation of \textit{HASP}, we prove that various ASR systems share significant similarity in low level feature extraction process.
\balance
\bibliographystyle{IEEEtran}
\bibliography{iccad_hasp}
\end{document}